\documentclass[aps,prl,twocolumn,groupaddress,longbibliography,footinbib]{revtex4-1}  %
\usepackage{graphicx} 
\usepackage{float} 
\usepackage{dcolumn} 
\usepackage{bm,color}
\usepackage{amssymb}
\usepackage{amsmath}
\usepackage{extarrows}
\usepackage[colorlinks=true,linkcolor=blue,urlcolor=blue,citecolor=blue]{hyperref}

\usepackage[normalem]{ulem} 
\usepackage{color}

\begin{document}
\title{Floquet engineering of optical solenoids and quantized charge pumping along tailored paths in two-dimensional Chern insulators}

\author{Botao Wang}
\email{botao@pks.mpg.de}
\author{F. Nur \"{U}nal}
\email{unal@pks.mpg.de}
\author{Andr\'{e} Eckardt}
\email{eckardt@pks.mpg.de}
\affiliation{Max-Planck-Institut f{\"u}r Physik komplexer Systeme, N{\"o}thnitzer Stra\ss e 38, 01187 Dresden, Germany}

\date{\today}

\begin{abstract}
The insertion of a local magnetic flux, as the one created by a thin solenoid, plays an important role in \emph{gedanken} experiments of quantum Hall physics. By combining Floquet engineering of artificial magnetic fields with the ability of single-site addressing in quantum gas microscopes, we propose a scheme for the realization of such local solenoid-type magnetic fields in optical lattices. We show that it can be employed to manipulate and probe elementary excitations of a topological Chern insulator. This includes quantized adiabatic charge pumping along tailored paths inside the bulk, as well as the controlled population of edge modes.
\end{abstract}

\maketitle

\paragraph{Introduction.}
The adiabatic creation of a single quasiparticle or quasihole in a quantum Hall insulator by inserting a magnetic flux quantum, using an infinitely thin solenoid, is a famous \emph{gedanken} experiment of quantum Hall physics \cite{1981Laughlin,2002Yoshioka}. Inspired by the recent experimental progress in controlling atomic quantum gases, here we propose and simulate a realistic scheme for the Floquet engineering of such strong solenoid-type local fluxes in an optical lattice system. The experimental realization of such an `optical solenoid' would provide a powerful novel tool for probing and manipulating topological states of matter and their excitations in a quantum gas.

In recent years, we have witnessed significant progress in the experimental realization and the coherent control of mesoscopic quantum systems of ultracold neutral atoms in optical lattice potentials. One milestone is the engineering of artificial magnetic fields \cite{2011Dalibard,2013Galitski,2014Goldman,2016Goldman,2017Eckardt,2009Lin,2011Aidelsburger,2012Struck,2013Struck,2013Aidelsburger,2015Kennedy,2015Aidelsburger,2017Tai,2014Atala,2015Mancini_edge,2015Stuhl,2017An,2014Jotzu,2017Tarnowski,2017Kolkowitz} which includes the creation of extremely strong magnetic fluxes of the order of one flux quantum per lattice plaquette (corresponding to hypothetical field strengths of $\sim 10^4$ Tesla for electrons in graphene) \cite{2009Lin,2011Aidelsburger,2012Struck,2013Struck,2013Aidelsburger,2015Kennedy,2015Aidelsburger,2017Tai}, the observation of topologically protected chiral edge transport \cite{2014Atala,2015Mancini_edge,2015Stuhl,2017An} and a bulk Hall response \cite{2014Jotzu,2015Aidelsburger}, as well as the measurement of non-zero Chern numbers \cite{2017Tarnowski,2015Aidelsburger}.
Furthermore, the ability of single-site addressing in quantum gas microscopes by using digital micromirror devices \cite{2016Ott_rev,2016Kuhr,2017Gross,2009Bakr,2010Bakr,2010Sherson,2015Parsons,2015Haller_Mic,2015Cheuk_Mic,2015Edge,2015Omran,2016Yamamoto, 2017Mitra,2012Cheneau,2015Islam,2016Kaufman,2017Choi,2016Greif,2017Mazurenko,2016Parsons,2016Boll,2016Cheuk,2017Hilker,2017Brown}  has been employed for studying the light-cone like spreading of correlations and information \cite{2012Cheneau}, measuring entanglement entropy \cite{2015Islam,2016Kaufman}, investigating many-body localization  \cite{2017Choi}, for the preparation of low-entropy states of matter \cite{2016Greif,2016Cheuk_FMI,2017Mazurenko}, and the detection of antiferromagnetic correlations \cite{2016Parsons,2016Boll,2016Cheuk,2017Hilker,2017Brown}.

\begin{figure}
	\centering\includegraphics[width=1\linewidth]{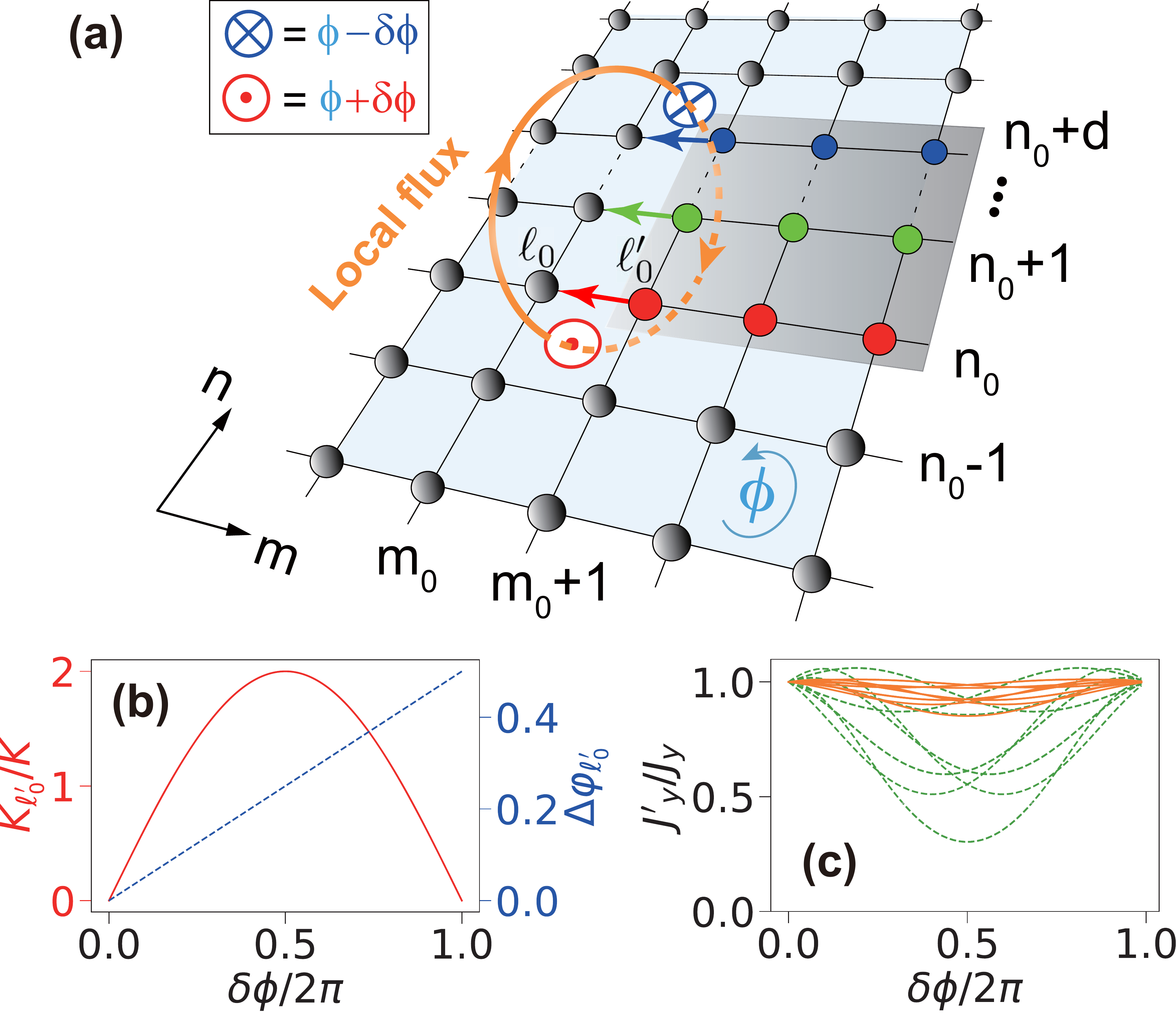}
	\caption{(a) Scheme for realizing solenoid-type local plaquette fluxes in a 2D optical lattice. The bonds labeled by colored arrows acquire an additional Peierls phase $\delta\phi$ as a result of the modification of the driving potentials on the colored rows. The integer $d$ represents the distance between the plaquettes labeled by $\otimes,\odot$ whose flux become $\phi\pm\delta\phi$,
with homogeneous background flux $\phi$.
	(b) Amplitude $K_{\ell_{0}^{\prime}}/K$ and phase $\Delta\varphi_{\ell_{0}^{\prime}}\equiv(\varphi_{\ell_{0}^{\prime}}-\varphi_{\ell_{0}^{\prime}\ell_{0}})/2\pi$ of the additional driving versus the phase shift $\delta\phi$ according to
Eq.~(\ref{Kvarphi}). (c) Modification of tunneling parameters transverse to the colored rows at the background flux $\phi=\pi/2$, for $K=0.15\hbar\omega$ (orange/solid) and $K=0.35\hbar\omega$ (green/dashed). }
	\label{fig_scheme}
\end{figure}

In this paper, we propose to exploit the single-site control and imaging provided by quantum gas microscopes for the realization of strong tunable artificial magnetic fields piercing single plaquettes of a two-dimensional (2D) optical lattice
[Fig.~\ref{fig_scheme}(a)]. We design a feasible scheme for the Floquet engineering of such optical solenoids based on photon-assisted tunneling induced by local potential modulations.
It can be used to adiabatically create and manipulate local quasiparticle/hole excitations in Chern insulator states. We show that in this way quantized charge pumping can be induced along tailored paths, which are determined by the artificial electric fields created by the time-dependent Peierls phases (vector potentials) used for the flux insertion.
This allows, among others, also for the coherent population (and depopulation) of edge modes, e.g.\ in order to probe chiral transport as a signature of the topological nature of the system.

\paragraph{Engineering a tunable local flux.}
Our starting point is the scheme for the realization of a homogeneous magnetic flux based on photon-assisted tunneling in a 2D optical square lattice \cite{2013Aidelsburger,2015Kennedy,2015Aidelsburger,2017Tai}.
We consider non-interacting particles subjected to an on-site potential $w_{\ell}\left(t\right)=w_{\ell}^{\text{dr}}\left(t\right)+\nu_{\ell}\hbar\omega$ on lattice site $\ell$.
Here $w_{\ell}^{\text{dr}}\left(t\right)=w_{\ell}^{\text{dr}}\left(t+T\right)$ denotes a time-periodic driving potential, which induces photon-assisted tunneling against the static energy offsets characterized by the integers $\nu_{\ell}$ and the angular driving frequency $\omega=2\pi/T$.
After a gauge transformation, the Hamiltonian takes the form \cite{2017Eckardt}
\begin{equation}
\hat{H}(t)=-{\displaystyle \sum_{\left\langle \ell, \ell^{\prime}\right\rangle }}J_{\ell^{\prime}\ell}e^{i\theta_{\ell^{\prime}\ell}(t)}\hat{a}_{\ell^{\prime}}^{\dagger}\hat{a}_{\ell},
\label{Ht}
\end{equation}
where $\hat{a}_\ell$~($\hat{a}^{\dagger}_\ell$) is the annihilation (creation) operator for a particle on site $\ell$ and $J_{\ell^{\prime}\ell}$ the parameter for tunneling between nearest-neighboring sites $\ell$ and $\ell^{\prime}$. The time-dependent Peierls phases $\theta_{\ell^{\prime}\ell}(t)=\int_{t_0}^{t}dt^{\prime}w_{\ell^{\prime}\ell}(t^{\prime})/\hbar+\chi^{(0)}_{\ell'}-\chi^{(0)}_{\ell}$ play the role of vector potentials, where the free gauge parameters $\chi^{(0)}_{\ell}$ are chosen to remove $t_0$-dependent terms.
For sinusoidal driving, the relative modulation between two neighboring sites $w_{\ell^{\prime}\ell}(t)=w_{\ell^{\prime}}(t)-w_{\ell}(t)$ takes the form of
\begin{equation}
w_{\ell^{\prime}\ell}(t)=K_{\ell^{\prime}\ell}\cos\left(\omega t-\varphi_{\ell^{\prime}\ell}\right)+(\nu_{\ell^{\prime}}-\nu_{\ell})\hbar\omega,
\label{w_ll}
\end{equation}
where $K_{\ell^{\prime}\ell}>0$ and $\varphi_{\ell^{\prime}\ell}$, respectively, represent the driving strength and phase.
In the high-frequency regime ($\hbar\omega\gg J_{\ell^{\prime}\ell}$), the time-dependent Hamiltonian (\ref{Ht}) can be approximated by its cycle average \cite{2017Eckardt}
 \begin{equation} \label{eq_H_eff}
\hat{H}^{\text{eff}}=\frac{1}{T}\intop_{0}^{T}dt\hat{H}(t)=-\sum_{\left\langle \ell,\ell^{\prime}\right\rangle }J_{\ell^{\prime}\ell}^{\text{eff}}\hat{a}_{\ell^{\prime}}^{\dagger}\hat{a}_{\ell}.
\end{equation}

Let us now consider a sinusoidally driven superlattice with staggered potential offsets in $x$-direction \cite{2015Aidelsburger}, $\nu_\ell=[1+(-1)^{m}]/2$. We use integer indices $m$ and $n$ to label the
$x$ and $y$ coordinates of each site $\ell=(m,n)$, as depicted in
Fig.~\ref{fig_scheme}(a). This results in effective tunneling parameters
\begin{align}
J_{x}^{\text{eff}}&\equiv\left.J_{\ell^{\prime}\ell}^{\text{eff}}\right|_{x}=J_{x}\mathcal{J}_{1}\left(\frac{K_{\ell^{\prime}\ell}}{\hbar\omega}\right)e^{i\theta_{\ell^{\prime}\ell}^{\text{eff}}},
\label{eq_Jx_ time averaged}\\
J_{y}^{\text{eff}}&\equiv\left.J_{\ell^{\prime}\ell}^{\text{eff}}\right|_{y}=J_{y}\mathcal{J}_{0}\left(\frac{K_{\ell^{\prime}\ell}}{\hbar\omega}\right). \label{eq_Jy_ time averaged}
\end{align}
Here $J_{x} (J_y)$ is the amplitude of bare tunneling along the $x$ ($y$)-direction, $\mathcal{J}_{\alpha}(\cdot)$ a Bessel function, and $\theta_{\ell^{\prime}\ell}^{\text{eff}}=\left(-1\right)^{m+1}\varphi_{\ell^{\prime}\ell}$ an effective time-independent Peierls phase.
Homogeneous tunneling amplitudes $\left|J_{x}^{\text{eff}}\right|$ and $\left|J_{y}^{\text{eff}}\right|$ ($K_{\ell^{\prime}\ell}=K$) and spatially varying Peierls phases of the form $\theta_{\ell^{\prime}\ell}^{\text{eff}}=\phi n + \eta_m $ can be achieved by oscillating site-dependent light-shift potentials \cite{2015Aidelsburger}.
This corresponds to a homogeneous magnetic flux of $\phi$ piercing each plaquette, not depending on the $m$-dependent term $\eta_m$.
In such a system, the Harper-Hofstadter model with $\phi=\pi/2$ has been successfully engineered experimentally  \cite{2015Aidelsburger}.

Starting from this configuration, we now describe how to engineer additional solenoid-type fluxes piercing two lattice plaquettes [Fig.~1(a)]. For this purpose, we consider additional driving potentials $K_{\ell}\sin\left(\omega t-\varphi_{\ell}\right)$ induced by digital mirror devices in the shaded subregion of Fig.~1(a). We first concentrate on tunneling along $x$-direction.
Within the rows labeled by the same color, identical driving is imposed on every site.
Thus, the tunneling processes within these rows remain unaffected.
However, for tunneling on the link connected by an unmodified site and a modified one [denoted by arrows in Fig.~\ref{fig_scheme}(a)], the relative modulation [Eq.~(\ref{w_ll})] is changed. We choose the additional driving so that the strength of tunneling remains the same, while the tunneling phase obtains a shift $\delta\phi$. The parameter for tunneling leftward along that link shall be modified to
\begin{equation}
J_{x}^{\prime\text{eff}}=J_{x}^{\text{eff}}e^{i\delta\phi}=\left|J_{x}^{\text{eff}}\right|e^{i\left(\phi n_0+\eta_{m_0}+\delta\phi\right)}.
\label{Jx_eff}
\end{equation}
In the $n_0$th row, e.g., this can be achieved by applying additional driving potential $K_{\ell_0'}\sin(\omega t-\varphi_{\ell'_0})$ on $\ell_0'$ (and all other red sites), with
\begin{equation}
K_{\ell_{0}^{\prime}}=-2K\sin\left(\delta\phi/2\right)
\text{ and }
\varphi_{\ell_{0}^{\prime}}=\varphi_{\ell_{0}^{\prime}\ell_{0}}-\delta\phi/2,
 \label{Kvarphi}
\end{equation}
The values of $\delta\phi$ can be varied continuously from 0 to 2$\pi$ by simultaneously tuning both the strength $K_{\ell_{0}^{\prime}}$ and the phase $\varphi_{\ell_{0}^{\prime}}$ of the additional driving as depicted in Fig.~\ref{fig_scheme}(b). For the other rows in the shadowed area, the same strategy is applied and the same $\delta\phi$ is implemented on each of the modified links [labeled by arrows in Fig.~\ref{fig_scheme}(a)].
This results in solenoid-type local fluxes $\phi\pm\delta\phi$ for the plaquettes at the end of the modified links [denoted by $\otimes,\odot$ in Fig.~\ref{fig_scheme}(a)], while the other plaquettes remain to have the uniform background flux $\phi$.

According to Eq.~(\ref{Kvarphi}), the additional driving phase $\varphi_{\ell^{\prime}}$ in the $n$th row depends on the original phase $\varphi_{\ell^{\prime}\ell}$ on the modified bond in that row.
Such row-dependent additional driving will modify the $y$ tunneling matrix elements in the modified area to
\begin{equation}
J_{y}^{\prime\text{eff}}=J_{y}\mathcal{J}_{0}\left(\frac{K\cdot c_{\ell'\ell}}{\hbar\omega}\right),
\label{Jy_eff}
\end{equation}
where the modified coefficients $c_{\ell'\ell}$ take finite bond-dependent
values.
In the case of a background flux $\phi=\pi/2$, the tunneling matrix elements are modified in eight different ways (depending on their position) that are
plotted in Fig.~\ref{fig_scheme}(c) for two different values of $K/(\hbar\omega)$.
Since the Bessel function $\mathcal{J}_{0}(x)\simeq 1-x^{2}/4$ changes only slowly
for small arguments, the so-induced tunnel inhomogeneities are reduced for small $K/\hbar\omega$ [Fig.~\ref{fig_scheme}(c)]. Moreover, since we are interested in (topologically non-trivial) insulating states, which are protected by a gap, the system will be robust against small inhomogeneities.

Ideally, the engineered system is captured by a modified Harper-Hofstadter model
\begin{equation}
\hat{H}^{\text{ideal}}=-{\displaystyle \sum_{\left\langle \ell,\ell^{\prime}\right\rangle }}Je^{i\left(\phi n+\mu_{\ell^{\prime}\ell}\delta\phi\right)}\hat{a}_{\ell^{\prime}}^{\dagger}\hat{a}_{\ell},
\label{Hideal}
\end{equation}
with a homogeneous tunneling amplitude $J$ and coefficients
$\mu_{\ell^{\prime}\ell}$ that take values 1 (-1) for tunneling along (against)
an arrow in Fig.~\ref{fig_scheme}(a) and 0 otherwise. For a quarter flux, the system possess four  bands. The lowest band, characterized by the topologically non-trivial Chern number $C=1$, shall be filled completely with noninteracting spinless fermions, so that the system forms an integer-quantum-Hall-type Chern insulator  \cite{2010Hasan_TI,2011Qi}.

\begin{figure}
\centering\includegraphics[width=1\linewidth]{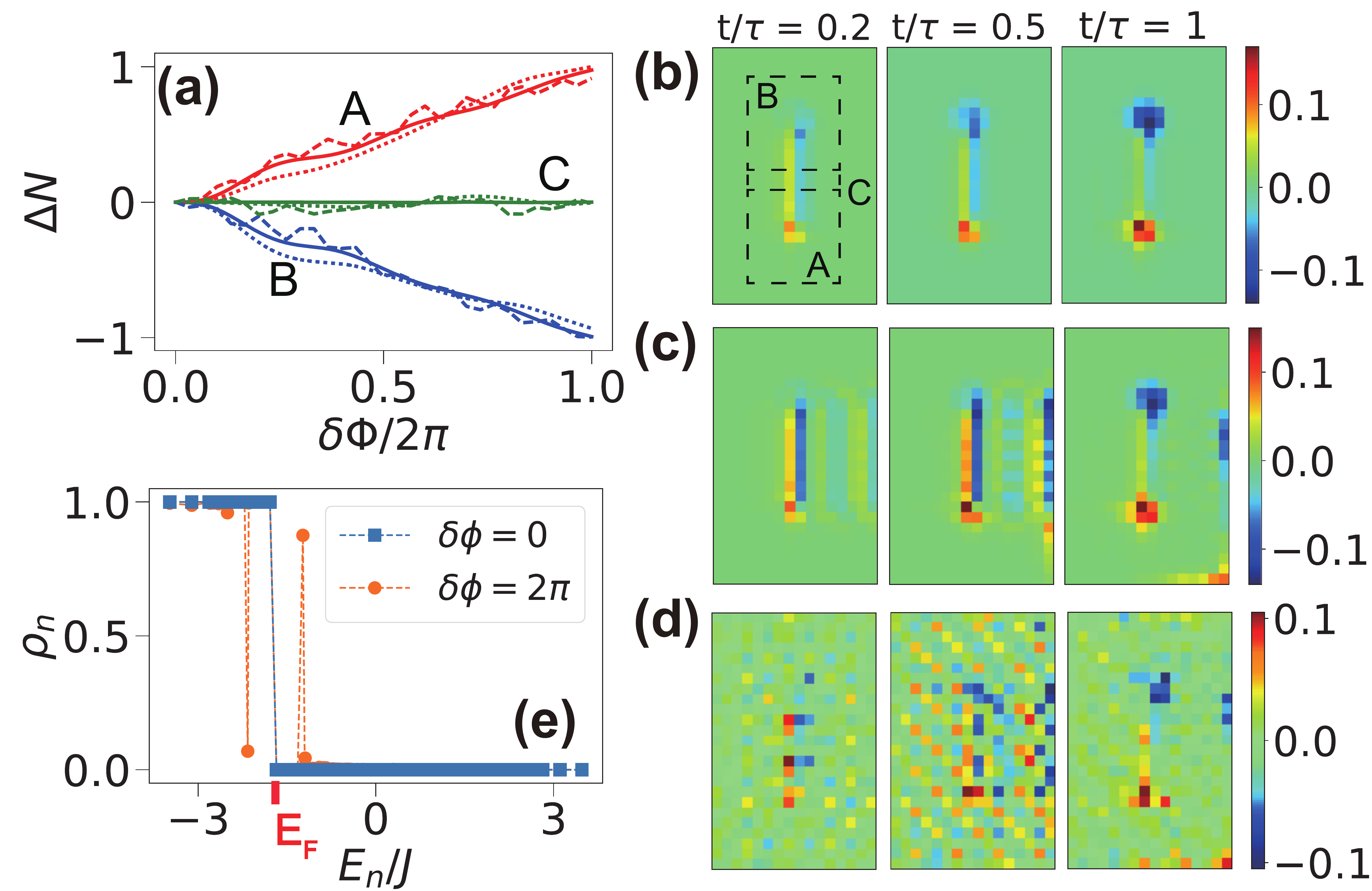}
\caption{(a) The particle number difference $\Delta N_{i} = N_{i}(t)-N_{i}(t=0)$
in regions $i=$ A, B and C [defined in (b)] as a function of $\delta\phi(t)$,
which is ramped linearly from $0$ to $2\pi$ within the time $\tau=10 \hbar/J$,
for $K/\hbar\omega=0.35$. The solid, dotted, and dashed lines are obtained by
solving the time-dependent Schr\"{o}dinger equation for the ideal Hamiltonian
(\ref{Hideal}), the time-averaged Hamiltonian (\ref{eq_H_eff}), and the
time-dependent Hamiltonian (\ref{Ht}) respectively. Snapshots of the respective
changes in spatial densities are depicted in (b),~(c)~and~(d). The simulation of
Hamiltonian (\ref{Ht}) includes the preparation of the Chern insulator before
the ramp (see Fig.~\ref{fig_exp}), and also imperfections as they would arise from a limited spatial resolution of the additional driving potentials.
(e) Population of single-particle energy eigenstates after the flux insertion computed for the ideal Hamiltonian (\ref{Hideal}). }
	\label{fig_pump}
\end{figure}

\paragraph{Creating quasiparticles and quasiholes.}
For $C=1$, we expect that locally a quasiparticle (quasihole) of ``charge" +1 (-1) can be created adiabatically by linearly increasing $\delta\phi$ from 0 to 2$\pi$ within a  proper ramping time $\tau$, i.e.\ $\delta\phi\left(t\right)=(2\pi/\tau) t$ \footnote{Strictly speaking, in a finite system a perfectly adiabatic dynamics would lead the system back to the ground state. However, the time scale of such a perfect adiabatic motion increases exponentially with the distance $d$ of the two plauqettes where the additional fluxes are inserted. For $d=10$ used here, the corresponding ramp time would be much longer than any typical experimental protocol.}. We consider a square lattice of size $16{\times}25$ with the two modified plaquettes separated by 10 lattice constants.
As a signature of such an excitation, we track the particle number differences $\Delta N_{i}(t)=N_{i}(t)-N_{i}(t=0)$ at time $t$ in three regions $i=A, B, C$.
We define that A (B) is a square-shaped region of size $10{\times}10$, centered at the plaquette with additional flux $\delta\phi$ $(-\delta\phi)$, and C is the area between A and B [see Fig.~\ref{fig_pump}(b)].
We numerically calculate the spatial density distribution by solving the time-dependent Schr\"{o}dinger equation for the ideal Hamiltonian (\ref{Hideal}) [Fig.~\ref{fig_pump}(b)], the
time-averaged Hamiltonian (\ref{eq_H_eff}) [Fig.~\ref{fig_pump}(c)], and the time-dependent Hamiltonian (\ref{Ht}) [Fig.~\ref{fig_pump}(d)]. In the latter case, we simulated also the
preparation of the Chern insulator state before ramping up $\delta\phi$, following the protocol described below [cf.\ Fig.~\ref{fig_exp}], where we also included imperfections mimicking a possible limited spatial resolution of the additional driving. For all three simulations, we plot $\Delta N_{i}$ as a function of
$\delta\phi$ [Fig.~\ref{fig_pump}(a)]. Good agreement among all these simulations
is observed, despite the fact that the simulation of the full time evolution
takes into account possible imperfections, non-adiabatic processes, and driving-induced heating.
The values $\Delta N_i$ near $\pm 1$ at $\delta\phi=2\pi$ suggest the creation of a quasiparticle (quasihole) around the plaquette where the additional flux $\delta\phi$$(-\delta\phi)$ was inserted. Note that static pinning potentials $V/J= \pm 1$ have been applied on each of the four sites around the plaquette with additional flux $\mp\delta\phi$, to prevent the created excitations from dispersing \cite{2015Liu}. This is necessary, because, unlike the Landau levels of a continuous system, the energy bands of the Harper-Hofstadter model are not perfectly flat. $V$ is small enough to cause only local density disturbances in the ground state [visible also in the plot for $t=\tau_2$ in Fig.~\ref{fig_exp}(b)]. Since the pinning potentials are present before the flux insertion, they do not contribute to the relative density change plotted in Fig.~\ref{fig_pump}.

In order to locate the quasiparticle and quasihole in the energy spectrum $E_n$, in Fig.~\ref{fig_pump}(e) we plot the mean occupation $\rho_n$ of the eigenstates of Hamiltonian (\ref{Hideal}) after the flux insertion. A quasihole appears in the occupied lowest band, while a quasiparticle is created in the first excited band.
Note that without pinning potentials, the quasiparticle and quasihole excitations are distributed over many states, while they correspond to eigenstates in the presence of the pinning potentials.

The snapshots of the evolution of the spatial density distributions are plotted in Figs.~\ref{fig_pump}(b-d), in which we can observe clear signatures of the creation of a quasiparticle and a quasihole. Figs.~\ref{fig_pump}(c),(d) show that the tunneling inhomogeneity [not present in the ideal model underlying Fig.~\ref{fig_pump}(b)] mainly induces extra excitations at the right edge. This is related to the fact that the edge modes are gapless. However, it does not affect the creation of quasiparticles and quasiholes in the bulk, which is protected by an energy gap [Fig.~\ref{fig_pump}(a)].

\begin{figure}
	\centering\includegraphics[width=1\linewidth]{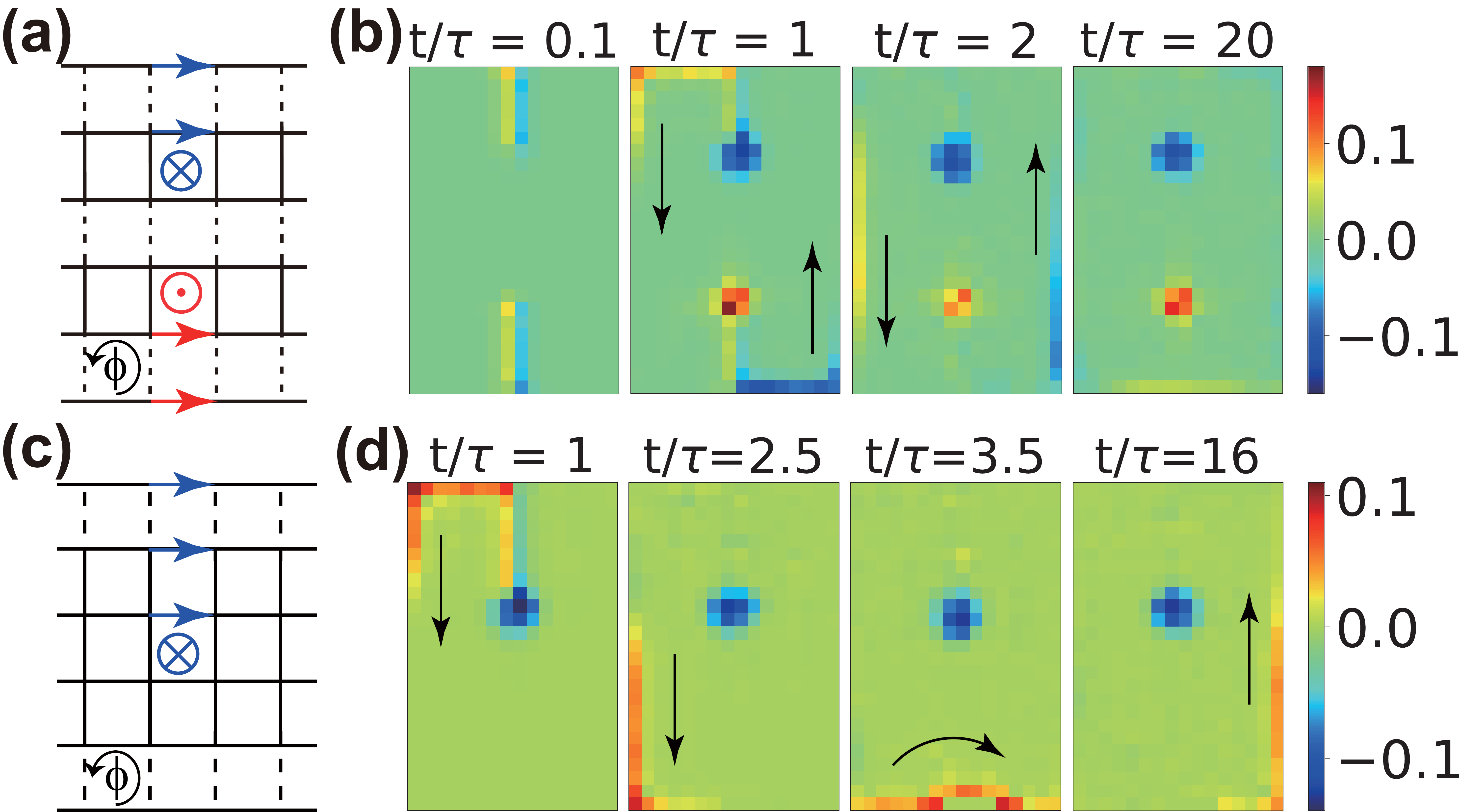}
	\caption{(a) Local flux configuration as in Fig.~\ref{fig_scheme}(a), but for
a different gauge. (b) Snapshots of corresponding spatial density
distributions during ramp.
(c) and (d) for additional flux in a single plaquette. The
chiral edge transport visible in (d) is robust at the corners and against an
impurity potential of $10J$ at the lower edge. The results are obtained from time
evolution governed by the ideal Hamiltonian (\ref{Hideal}) with $\delta\phi(t)=2\pi\tau/t$ for $t<\tau$ and $\delta\phi(t)=2\pi$ for $t>\tau$, where
$\tau=10\hbar/J$. }
	\label{fig_edge}
\end{figure}

\paragraph{Controlled charge pumping.}
Fig.~\ref{fig_pump}(b) shows that a quantum of charge is transported along the path defined by the sequence of bonds with the additional Peierls phase $\delta\phi$. As shown in Fig.~\ref{fig_edge}(a), by adding phases $\delta\phi$ on the bonds starting from the target plaquettes to the edge instead of between them, a different particle transport is induced. Apart from the previous bulk excitations, a quasiparticle and a quasihole appear at the edges [Fig.~\ref{fig_edge}(b)]. Again quanta of charge are transported along the path defined by the modified Peierls phases. After being created, such edge excitations follow a chiral motion, until their densities annihilate each other after some time.
By introducing a flux through a single plaquette [Fig.~\ref{fig_edge}(c)], the edge mode can be coherently populated in a controlled fashion along the path defined by the additional Peierls phases [Fig.~\ref{fig_edge}(d)].
This mechanism can be a useful tool to probe robust chiral edge transport as visualized in Fig.~\ref{fig_edge}(d).

In our 2D system, the path along which a quantum of charge is transported can be controlled by choosing a particular gauge for the additional flux. This can be understood by noting that artificial electric fields $\propto\dot{\theta}_{\ell^{\prime}\ell}$ are induced when changing the Peierls phases $\theta_{\ell^{\prime}\ell}$ between sites $\ell^{\prime}$ and $\ell$ (which represent a vector potential).
Thus, the optical solenoid fields proposed here allow for quantized charge pumping along tailored paths in 2D topological Chern insulators.

\paragraph{Experimental protocol.}
For the implementation of the above-mentioned local flux scheme, we consider  the following experimental protocol, which includes also the preparation of the Chern insulator state from the ground state of the undriven lattice [Fig.~\ref{fig_exp}(a)].
We start with a trivial band insulator in a staggered potential with energy offsets $\Delta+\delta$ along $x$ and $\delta$ along $y$ direction \cite{2015Aidelsburger}.
Then the driving amplitude $K$ is ramped up linearly to its final value within time $\tau_{1}$.
Subsequently the detuning $\delta/J$  is ramped down between $\tau_{1}$ and $\tau_{2}$. During this ramp, the system undergoes a topological phase transition into a Chern insulator state \cite{2015Aidelsburger}. This topologically non-trivial regime should be reached adiabatically by relying on the finite extent of the system.
At the same time, the pinning potentials ($|V|/J=1$) are also switched on at the modified plaquettes. In the last stage ($\tau_{2}\rightarrow\tau_{3}$), the additional flux $\delta\phi$ is ramped from 0 to $2\pi$ by tuning the amplitudes $K_{\ell}^{\prime}$ and the phases $\varphi_{\ell}^{\prime}$ according to Eq.~(\ref{Kvarphi}). Furthermore,
to address any concern regarding possible experimental imperfections of the single-site resolution, we consider only 80\% of the proposed ideal additional driving on site $\ell$ to act
on this site, while the remaining 20\% are distributed evenly among the four neighboring sites. The results obtained from integrating the time evolution for this protocol using the time-dependent Hamiltonian (\ref{Ht}) are presented in Fig.~\ref{fig_pump}(a) (dashed lines) and (d). Since our simulation starts with adiabatic preparation from a trivial band insulator, possible driving induced interband transitions \cite{ThomasInterband} are also included.
We observe that the quantized charge transport is spoiled neither by such interband excitations nor by the imperfect spatial resolution [Fig.~\ref{fig_pump}(a),~(d)]. Note that not taking into account interactions is well justified in systems of spin-polarized fermionic atoms, and that the heating in Floquet systems has not been a major problem in experiments with such spin-polarized fermions \cite{2014Jotzu}.

\begin{figure}
	\centering\includegraphics[width=1\linewidth]{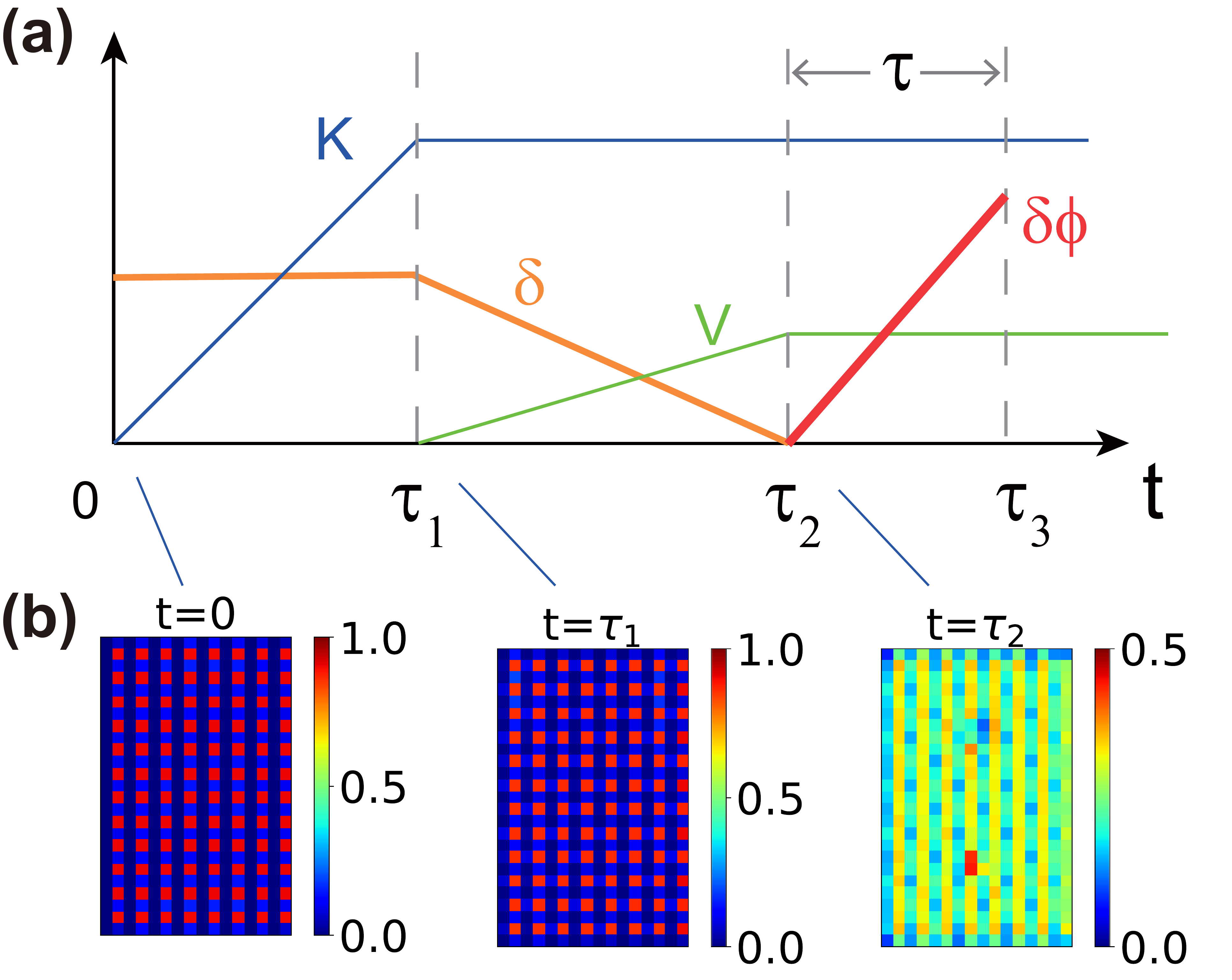}
	\caption{(a) Experimental protocol for the preparation of the Chern insulator starting from the ground state of the undriven lattice and the subsequent flux insertion (see main text).
(b) Spatial density distribution at different stages of the protocol from the simulation of the time evolution of Eq.~(\ref{Ht}), for driving frequency of $\hbar\omega=70J$, initial superlattice strength $\delta=4J$, final driving amplitude
$K/\hbar\omega=0.35$, as well as ramping times $\tau_{1}=200\hbar/J,
\tau_{2}-\tau_{1}=400\hbar/J,\tau=\tau_{3}-\tau_{2}=10\hbar/J$. The effect of the flux insertion ($\tau_2 \rightarrow \tau_3$) is shown in Fig.~\ref{fig_pump}(d).}
	\label{fig_exp}
\end{figure}

\paragraph{Conclusion and outlook}
We have proposed a scheme for the realization of tunable solenoid-type local magnetic fluxes in 2D optical lattices. Such optical solenoids allows for quantized charge pumping in a topological Chern insulator along tailored paths. This effect can, for example, be employed for the coherent population of chiral edge modes to probe their robust chiral transport properties. An interesting application for optical solenoids can also be to create, probe and manipulate fractionally charged excitations of fractional Chern insulators \cite{2018Raviunas}.

\begin{acknowledgments}
	\emph{Acknowledgment.}
	We thank Egidijus Anisimovas and Mantas Ra\v{c}i\={u}nas for helpful discussions, and acknowledge the support from the Deutsche Forschungsgemeinschaft (DFG) via the Research Unit FOR 2414.
\end{acknowledgments}

\bibliography{LocalFlux}

\end{document}